\documentclass[aip,pop,reprint,numerical,twocolumn]{revtex4}
\usepackage{amssymb, amsmath,framed}
\usepackage{color,soul}

\usepackage{txfonts}               
\usepackage{graphicx}
\usepackage{dcolumn}
\usepackage{bm}
\usepackage{color}
\usepackage{mathrsfs}

\usepackage{bm}
\usepackage[colorlinks=true,linkcolor=blue]{hyperref}
\expandafter\ifx\csname package@font\endcsname\relax\else
 \expandafter\expandafter
 \expandafter\usepackage
 \expandafter\expandafter
 \expandafter{\csname package@font\endcsname}
\fi
\def \mr#1{\mathrm{#1}}

\usepackage{mathrsfs}
\DeclareMathAlphabet{\mathpzc}{OT1}{pzc}{m}{it}

\def\rme{\mathrm{e}}

\def\rmH{\mathrm{H}}

\begin{document}

\title{Anisotropy in broad component of H$\alpha$ line in the magnetospheric device RT-1}

\author{Y. Kawazura}
\email{kawazura@ppl.k.u-tokyo.ac.jp}
\author{N. Takahashi, Z. Yoshida, M. Nishiura, T. Nogami, A. Kashyap, Y. Yano, H. Saitoh, M. Yamasaki, T. Mushiake, and M. Nakatsuka}
\affiliation{
Graduate School of Frontier Sciences, The University of Tokyo, 5-1-5 Kashiwanoha, Kashiwa 277-8561, Japan
}


\begin{abstract}
Temperature anisotropy in broad component of H$\alpha$ line was found in the ring trap 1 (RT-1) device by Doppler spectroscopy.
Since hot hydrogen neutrals emitting a broad component are mainly produced by charge exchange between neutrals and protons, the anisotropy in the broad component is the evidence of proton temperature anisotropy generated by betatron acceleration. 
\end{abstract}

\maketitle


%
%
\section{Introduction}
Magnetospheres exhibit various intriguing phenomena.
A representative example is plasma density profile with planet-ward gradient~\cite{Schulz}.
Such a self-organized structure is an apparent counterexample of the entropy principle.
The inward diffusion theory has successfully explained the self-organization mechanism in a magnetosphere;
diffusion triggered by violation of a third adiabatic invariant flattens their density on magnetic coordinates rather than Cartesian coordinates\,\cite{Hasegawa,YM2014,Sato2015-1}, resulting in nonuniform density distribution on Cartesian coordinates.
Such a nonuniform density profile was experimentally observed in the ring trap 1 (RT-1)~\cite{Yoshida2010,Saitoh2014} and Levitated Dipole Experiment (LDX)~\cite{Boxer}.

Another interesting self-organization phenomenon in magnetosphere is plasma heating by inward diffusion.
Particles are energized along with the inward (i.e., toward strong magnetic field region) transport keeping first and second adiabatic invariants constant. 
The conservation of the first adiabatic invariant increases the particle's perpendicular kinetic energy (betatron acceleration) and the conservation of the second results in the increase in parallel energy (Fermi acceleration)~\cite{Dessler}. 
These are the primary generation mechanism for planetary radiation belts.
In point dipole field configuration, betatron acceleration is stronger than Fermi acceleration because increase rate of magnetic field strength upon the planet-ward displacement is greater than the decrease rate of longitudinal bounce length. 
Therefore accelerated particle exhibits temperature anisotropy ($T_\perp > T_{||}$).
Recently we observed ion temperature anisotropy in RT-1 and proved that the anisotropy was produced by the betatron acceleration~\cite{Kawazura2015}. 

In this study, we found similar temperature anisotropy in the broad component of H$\alpha$ line by Doppler spectroscopy.
Broad component of H$\alpha$ line has been observed in astrophysical and laboratory plasmas.
In astrophysics, Balmer lines emitted around shocks in supernova remnants show clear narrow and broad components. 
The broad component is attributed to hot hydrogen particles produced through charge exchange between post-shock protons.
The narrow component originates in cold hydrogen atoms in pre-shock interstellar medium~\cite{broad-in-shock1,broad-in-shock2}.
The post shock proton temperature is estimated by Doppler broadening of the broad component.
Broad component of Balmer lines has also been observed in laboratory plasma~\cite{Kasai,Samm,Kubo}.
In addition to narrow and broad components, the recent studies identified threefold (cold, warm and hot) Balmer lines~\cite{Iwamae, Shikama, Fujii2013}.
The one-dimensional neutral particle transport model clarified that the hot component was generated by charge exchange with protons~\cite{Fujii2013}.
Unlike in the case of astrophysical plasma, it is challenging to infer proton temperature via Doppler temperature of a broad component in laboratory.
The preceding studies observed discrepancies between the proton temperature and the Doppler temperature of the broad component~\cite{Kasai, Kubo}.
The discrepancy originates in other processes which generate hot neutrals, such as dissociation and reflection.
However, the recent study succeeded to estimate proton temperature from H$\alpha$ line shape by separating charge exchange contribution from other competitive processes~\cite{Wan}.
Therefore, our observation of the temperature anisotropy in the broad component has the potential to estimate proton temperature anisotropy.

%
%
\section{Experimental set-up}
\begin{figure*}[htpb]
	\begin{center}
		\includegraphics*[width=0.8\textwidth]{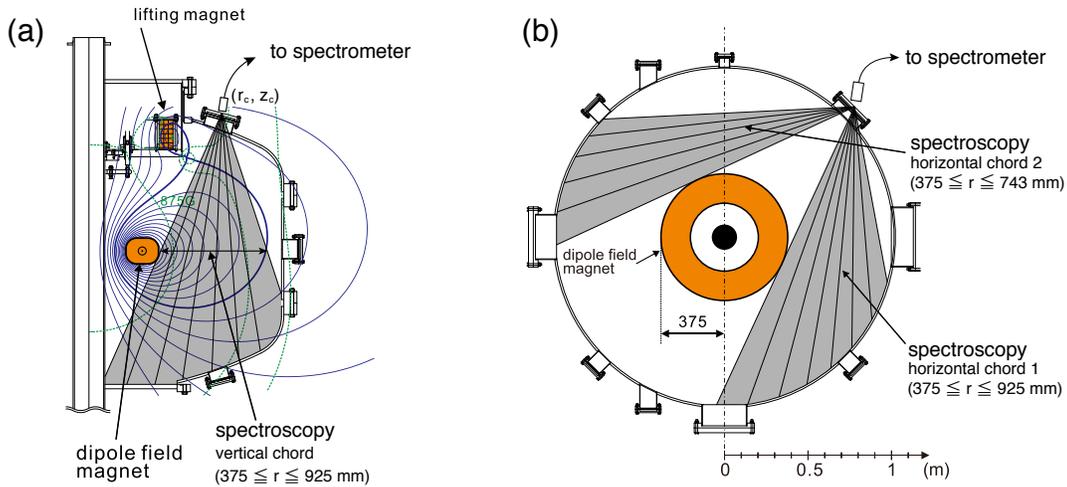}
	\end{center}
	\caption{ (a) The vertical and (b) horizontal cross-sectional views of RT-1 device. 
					}
	\label{f:layout}
\end{figure*}
The experiment was performed on RT-1 device which simulates a `laboratory magnetosphere' by a levitated superconducting magnet~\cite{Yoshida2006}.
Plasma is generated by the electron cyclotron resonance heating (ECH) using an 8.2\,GHz microwave.
The discharge duration is about 1\,s.
The electron density distribution (measured by three chord interferometers) has a peak near the dipole magnet and decreases toward outer wall~\cite{Saitoh2014,Saitoh2015}.
The electron density at the peak is on the order of $10^{17}\sim10^{18}\,\mr{m^{-3}}$. 
The hot electron temperature is 10$\sim$50\,keV and the cold electron temperature is 50$\sim$100\,eV.
The population of hot electron is approximately the same as that of cold electrons.
The heating mechanisms of ions are the thermal equilibration with cold electrons and the betatron acceleration, and the cooling mechanism is charge exchange with neutrals~\cite{Kawazura2015}.

\begin{figure}[htpb]
	\begin{center}
		\includegraphics*[width=0.5\textwidth]{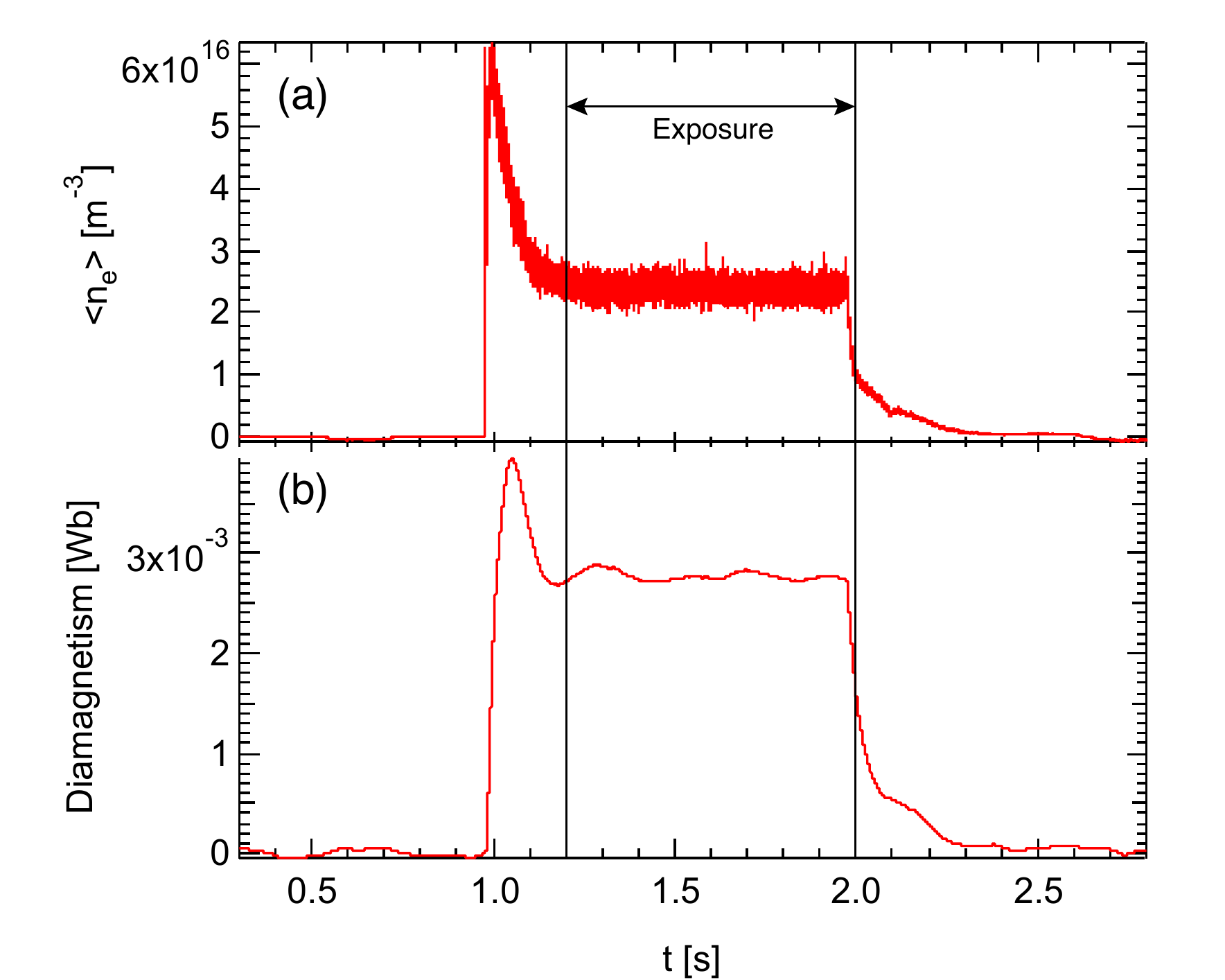}
	\end{center}
	\caption{
					Typical discharge waveforms of (a) line averaged density at $r=620$\,mm and (b) diamagnetic signal.
					The filling gas pressure and the ECH power for this shot were $5.4\times10^{-3}$\,Pa and 36\,kW.
					}
	\label{f:waveforms}
\end{figure}

We measured visible spectra using a Czerny-–Turner spectrometer (1\,m focal length, 2400\,grooves/mm grating, and 0.55\,nm/mm reciprocal linear dispersion) equipped with a charge-coupled device (CCD) detector (1024 $\times$ 256 pixels).
The wavelength resolution was 0.01\,nm.
Figure~\ref{f:layout} shows the cross-sectional views of RT-1. 
The line of sights for the three sets of Doppler spectroscopies ((a) one vertical chord and (b) two horizontal chords) are shown.
The vertical chord consisted of three fixed channels (548\,mm, 840\,mm, and 925\,mm) and one rotating channel which scanned $375\le r \le 925$\,mm (``vertical chord'' in Fig.~\ref{f:layout}(a)). 
The horizontal chord consisted of three fixed channels (772\,mm, 841\,mm, and 925\,mm) and two rotating channels.
One of the horizontally rotating channel scanned $375\le r \le 925$\,mm in the same direction as the three fixed channels (``horizontal chord 1'' in Fig.~\ref{f:layout}(b)) and the other rotating channel scanned $375\le r \le 743$\,mm in the symmetric direction (``horizontal chord 2'' in Fig.~\ref{f:layout}(b)).
For fixed channels, 10--11 shot data were averaged to improve signal-to-noise ratio whereas single shot data were acquire at each radial point for rotating channels.
The absolute value of the Doppler shift was estimated by combination of horizontal chord 1 and 2:
for a given radial point, unshifted central wavelength is obtained by midpoint of spectra measured by chord 1 and 2 assuming axisymmetric property.
Since RT-1 does not have toroidal magnetic field, the Doppler broadening measured in the horizontal chords gives temperature perpendicular to the ambient magnetic field ($T_\perp$).
On the other hand the vertical chord measures the mixture of $T_\perp$ and the parallel temperature ($T_{||}$).
Figure~\ref{f:waveforms} illustrates the typical discharge waveforms. 
The exposure time of spectrometer was 1.2--2.0\,s, which corresponded to the stable state of the discharge.

%
%
\section{Results}
\begin{figure}[htpb]
	\begin{center}
		\includegraphics*[width=0.5\textwidth]{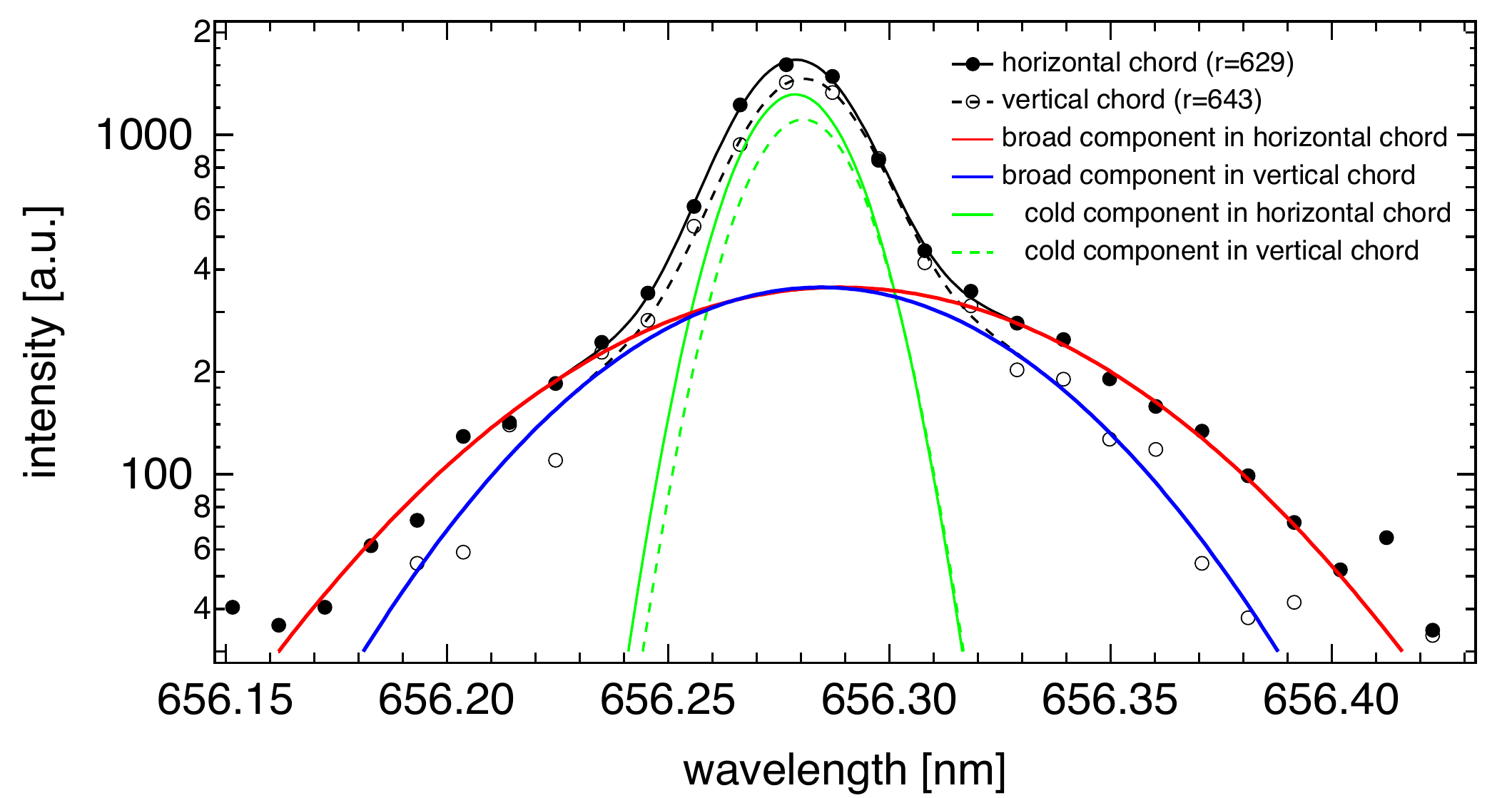}
	\end{center}
	\caption{
					Spectra of H$\alpha$ line measured by horizontal and vertical chords at almost the same radial point ($r\sim600$\,mm) and their dual Gaussian fittings.
					The vertical spectrum is normalized so that the broad component maximum coincides with that of horizontal chord.
					The filling gas pressure and the ECH power for this shot were $6.0\times10^{-3}$\,Pa and 35\,kW.
					}
	\label{f:dual_gaussian}
\end{figure}

Figure~\ref{f:dual_gaussian} shows the typical spectra of H$\alpha$ line in hydrogen discharge measured by the horizontal and vertical chords at almost the same radial point ($r\sim600$\,mm). 
The broad component was apparently seen.
Dual Gaussian fitting separates the narrow and the broad components.
The Doppler shift was finite only for the broad component in horizontal chord.
The broad component in horizontal chord was broader than that in vertical chord, showing the temperature anisotropy ($T_\perp > T_{||}$).
In what follows, we call the hot and cold neutrals as the emission sources of broad and narrow components respectively. 

\begin{figure}[htpb]
	\begin{center}
		\includegraphics*[width=0.5\textwidth]{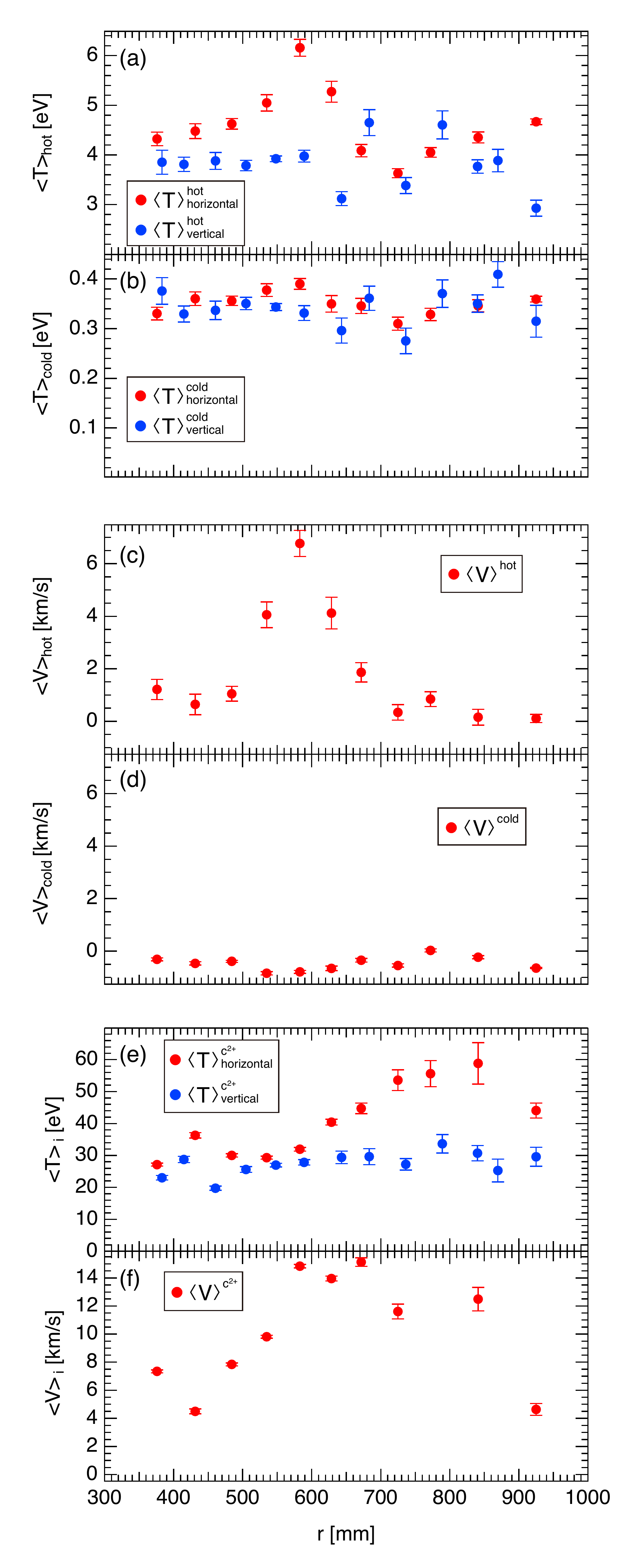}
	\end{center}
	\caption{
					Spatial profile of (a) hot and (b) cold neutral temperature, (c) hot and (d) cold neutral flow velocity, (e) temperature and (f) flow velocity of C$^{2+}$ ions.
					All the values are line averaged.
					The filling gas pressure and the ECH power for this shot were $5.4\times10^{-3}$\,Pa and 36\,kW.
					}
	\label{f:profile}
\end{figure}

Figure~\ref{f:profile} (a) illustrates the spatial profile of line averaged temperature of the hot neutrals.
$T_\perp$ of the hot neutrals had a peak in the middle of the confinement region ($r \sim 600$\,mm), then decreased towards dipole field magnet and the outer wall.
$T_{||}$ was, on the other hand, spatially homogeneous.
Figure~\ref{f:profile} (b) shows the temperature profile of the cold neutrals.
In contrast to the hot neutrals, cold neutral temperature was spatially homogeneous without temperature anisotropy.
Therefore, the temperature anisotropy was maximized at $r \sim 600$\,mm.
Figure~\ref{f:profile} (c) and (d) show the spatial profiles of line averaged toroidal flow velocity of the hot and cold components respectively.
The hot neutral flow speed had maximum at the same point of $T_\perp$'s maximum and decreased toward the both boundaries.
The cold neutrals, on the other hand, did not have toroidal flow.

For the comparison of the hot neutrals and ions, we present the spatial profiles of the temperatures and the toloidal flow speed of C$^{2+}$ impurity ions measured by C\,III (464.7\,nm) line in Fig.~\ref{f:profile} (e) and (f), respectively.
The ion temperature had maximum at $r\sim800$\,mm. The ion flow speed had maximum at $600\le r \le 700$\,mm.
Thus, spatial profiles are different between hot neutrals and impurity ions.
The absolute value of C$^{2+}$ temperature was ten times higher than that of hot neutrals, and flow speed is twice faster.
As we showed in the previous paper~\cite{Kawazura2015}, He$^+$ ions in helium discharge have similar temperature profile of C$^{2+}$ ions in Fig.~\ref{f:profile} (e). 
The discrepancy between hot neutrals and ions suggests that hot neutrals are not identical to protons as proven in the past studies~\cite{Kasai, Kubo}.

\begin{figure}[htpb]
	\begin{center}
		\includegraphics*[width=0.5\textwidth]{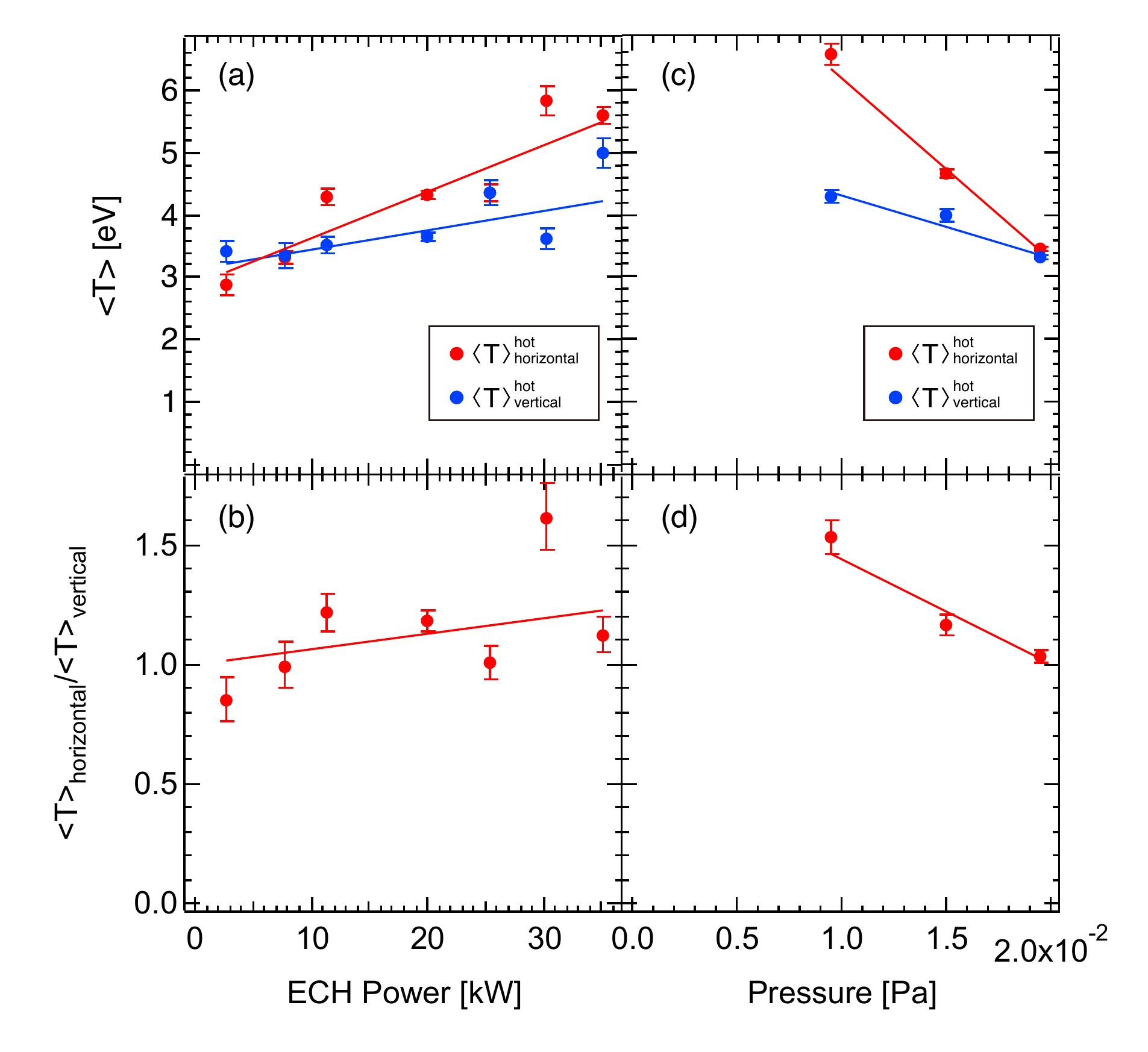}
	\end{center}
	\caption{
						Dependence of
						(a) the line averaged temperature and (b) the anisotropy of hot neutrals on the ECH power (filling gas pressure was fixed as $5\times10^{-3}$\,Pa),
						and the dependence of (c) the line averaged temperature and (d) the anisotropy on the filling gas pressure (ECH power was fixed as 35\,kW).
					}
	\label{f:dependence}
\end{figure}

Figure\,\ref{f:dependence} shows the dependencies of the line averaged hot neutral temperatures and anisotropy at $r=$583\,mm on ECH power and filling gas pressure.
The dipole field magnet was mechanically supported (i.e., without levitation) for this measurement.
From Fig.\,\ref{f:dependence}(a) and \ref{f:dependence}(b), both the temperatures and anisotropy increased as ECH power increased.
From Fig.\,\ref{f:dependence}(c) and \ref{f:dependence}(d) both the temperatures and anisotropy decreased as the filling gas pressure increased.
This tendencies coincided with those of ions~\cite{Kawazura2015}. 
For ions, the increase in ECH power results in increase in temperature and anisotropy through the relaxation with electrons because the betatron acceleration heating is proportional to $T_\perp$ itself. 
On the other hand the increase in the filling gas pressure leads decrease in the temperature and anisotropy through the charge exchange loss.

%
%
\section{Summary and Discussion}
We observed the broad component of H$\alpha$ line emission (hot neutrals).
The radial profile of the temperature and the toroidal velocity of hot neutrals were spatially inhomogeneous, whereas the cold component had flat profiles.
Furthermore, the hot neutral temperature had anisotropy ($T_\perp > T_{||}$). 
Even though the preceding studies already observed the hot neutral flow~\cite{Iwamae, Shikama, Fujii2013}, the observation of the temperature anisotropy in hot neutrals is the very first.
The dependencies of hot neutral temperature and anisotropy were consistent with ions.
Since hot neutrals emitting a broad component are mainly produced by charge exchange between neutrals and protons~\cite{Fujii2013}, the flow and the anisotropy were inherited from protons.
Therefore, our observation is the evidence of proton temperature anisotropy in RT-1.
Although the discrepancy between hot neutrals and impurity ions suggests that hot neutrals are not identical to protons, anisotropy in broad component may be a technique to estimate proton temperature anisotropy with help from precise analysis of line shape because the preceding study proved that separation of charge exchange contribution from competitive processes enables us to evaluate proton temperature~\cite{Wan}.

Now, let us discuss the locality of the present observation.
We may consider three scenarios for H$\alpha$ emission through charge exchange. 
\begin{eqnarray}
	&&\rmH_{(n=3)} + \rmH^+ \to \rmH^+ + \rmH_{(n=3)}
\label{e:3->3}
\end{eqnarray}
\begin{eqnarray}
	&&\rmH_{(n=2)} + \rmH^+ \to \rmH^+ + \rmH_{(n=2)} \nonumber\\ 
	&&\quad \quad \quad \quad \Rightarrow \quad \rmH_{(n=2)} + \rme^- \to \rmH_{(n=3)} + \rme^-
\label{e:2->3}
\end{eqnarray}
\begin{eqnarray}
	&&\rmH_{(n=1)} + \rmH^+ \to \rmH^+ + \rmH_{(n=1)} \nonumber\\
	&&\quad \quad \quad \quad \Rightarrow \quad \rmH_{(n=1)} + \rme^- \to \rmH_{(n=3)} + \rme^- 
\label{e:1->3}
\end{eqnarray}
Here, we consider only resonance charge exchange processes because cross section of non-resonance charge exchange is much smaller.
The process (\ref{e:3->3}) is resonance charge exchange of $n=3$ level.
Then, spontaneous emission of H$\alpha$ de-excites the charge exchanged atoms to $n=2$ level.
Since the spontaneous emission of $n=3\to2$ occurs in $\sim 2.3\times10^{-8}$\, s, the travel distance of hot neutral with 6\,eV from charge exchange to spontaneous emission is 0.5\,mm, implying the sufficient locality for this scenario.
The process (\ref{e:2->3}) is resonance charge exchange of $n=2$ level followed by electron impact excitation to $n=3$ level.
Then, spontaneous emission of H$\alpha$ will occur.
However, this scenario is not plausible because the electron impact excitation time (calculated as $5\times10^{-6}$\,s from $n_\mr{e}\sim 10^{17}\,\mr{m^{-3}}$ and $10\le T_\mr{e}\le 100$\,eV~\cite{IAEA-AMDIS}) is significantly shorter than the time of the spontaneous emission $n=2\to1$ ($\sim 2.1\times10^{-9}$\, s).
Therefore, almost all the charge exchanged atoms in $n=2$ level go down to the ground level instantly. 
The process (\ref{e:1->3}) is charge exchange of ground level followed by electron impact excitation to $n=3$ level.
The cross section of the resonance charge exchange in $n=1$ level is approximately ten times smaller than that of the resonance charge exchange in $n=3$ level~\cite{Fujii2011}.
The time of the electron impact excitation $n=1\to3$ is $2\times10^{-3}$\,s~\cite{IAEA-AMDIS}.
The travel distance of hot neutral with 6\,eV is 50\,m, which is incomparably larger than the device size.
Also, the cross section of electron impact ionization ($\rmH_{(n=1)} + \rme^- \to \rmH^+ + 2\rme^- $) is approximately ten times larger than that of excitation ($\rmH_{(n=1)} + \rme^- \to \rmH_{(n=3)} + \rme^-$)~\cite{IAEA-AMDIS}, and thereby the charge exchanged neutrals at $n=1$ level will be ionized before excitation to $n=3$ level.
These estimates suggests that the process (\ref{e:1->3}) is rare for a single neutral particle.
However, since the number of the ground state neutrals is much larger than that of $n=3$ state neutrals before charge exchange, the contribution of the process (\ref{e:1->3}) is not negligible.
Therefore, both of the scenarios (\ref{e:3->3}) and (\ref{e:1->3}) are feasible, and we have yet to determine which of the processes is governing.
Because the spatial resolution of the process (\ref{e:1->3}) is equivalent to the device size, the contribution of this process is, at most, only ``uniform background'' temperature in Fig.~\ref{f:profile}(a).
The peaks of Fig.~\ref{f:profile} (a) and (c) are attributed to the process (\ref{e:3->3}), and the spatial resolution is $\sim0.5$\,mm near the peaks.
Precise numerical simulation of neutral particle transport is necessary for further discussions.

In closing, we remark the comparison with tokamak configuration.
The preceding experiment in tokamak configuration measured the Doppler temperature of broad component from multiple directions which are tangential and perpendicular to toroidal magnetic field~\cite{Kasai} in the similar manner as the present study.
The observed broad component in H$\alpha$ was precisely isotropic. 
Therefore, an anisotropy in broad component is distinctive to magnetospheric configuration.

\section*{acknowledgments}
This work was supported by JSPS KAKENHI Grant No. 23224014.

\end{document}